\documentclass{sig-alternate-05-2015}
\usepackage{tabu}
\usepackage{algorithm}  
\usepackage{algorithmic}  
\usepackage{booktabs}
\begin{document}

\setcopyright{acmcopyright}

\doi{10.475/123_4}

\isbn{123-4567-24-567/08/06}

\conferenceinfo{WWW 2017}{April 3--7, 2017, Perth, Australia}

\acmPrice{\$15.00}

%

\title{User Personalized Satisfaction Prediction via\\
 Multiple Instance Deep Learning }

%
%
%
%
%

\numberofauthors{5} 
%
\author{
%
%
\alignauthor
Zheqian Chen\\
       \affaddr{State Key Lab of CAD \& CG}\\
       \affaddr{Zhejiang University}\\
       \email{zheqianchen@gmail.com}
\alignauthor
Ben Gao\\
       \affaddr{State Key Lab of CAD \& CG}\\
       \affaddr{Zhejiang University}\\
       \email{beng.zju@gmail.com}
\alignauthor 
Huimin Zhang\\
       \affaddr{State Key Lab of CAD \& CG}\\
       \affaddr{Zhejiang University}\\
       \email{rachelbowlong@gmail.com}
\and  
\alignauthor 
Zhou Zhao\\
       \affaddr{Key Lab of DCD}\\
       \affaddr{Zhejiang University}\\
       \email{zhaozhou@zju.edu.cn}
\alignauthor 
Deng Cai\\
       \affaddr{State Key Lab of CAD \& CG}\\
       \affaddr{Zhejiang University}\\
       \email{dengcai@gmail.com}      
}



\maketitle

\begin{abstract}
Community-based question answering(CQA) services have arisen as a popular knowledge sharing pattern for netizens. With abundant interactions among users, individuals are capable of obtaining satisfactory information. However, it is not effective for users to attain answers within minutes. Users have to check the progress over time until the satisfying answers submitted. We address this problem as a user personalized satisfaction prediction task. Existing methods usually exploit manual feature selection. It is not desirable as it requires careful design and is labor intensive. In this paper,  we settle this issue by developing a new multiple instance deep learning framework. Specifically, in our settings, each question follows a weakly supervised learning (multiple instance learning) assumption, where its obtained answers can be regarded as instance sets and we define the question resolved with at least one satisfactory answer. We thus design an efficient framework exploiting multiple instance learning property with deep learning tactic to model the question-answer pairs relevance and rank the asker's satisfaction possibility. Extensive experiments on large-scale datasets from Stack Exchange demonstrate the feasibility of our proposed framework in predicting askers personalized satisfaction. Our framework can be extended to numerous applications such as UI satisfaction Prediction, multi-armed bandit problem, expert finding and so on.  
\end{abstract}

\keywords{User Satisfaction Prediction; Multiple Instance Learning; Deep Learning}

\section{Introduction}
Community-based question answering(CQA) services have emerged as prevalent and helpful platforms to share knowledge and to seek information for netizens. With abundant interactions and fully openness, CQA services enable users to directly obtain specific information from other community participants. However, there is a fundamental problem in CQA services. Users may take days or even weeks to wait for a satisfactory answer posted. It is too time-consuming and users may not have so much patience to check the progress and get the question resolved. Hence how to predict the user's personalized satisfaction have become inevitably crucial. In this paper, we target at predicting the user's individual satisfaction possibility. It is meaningful to resolve this challenge, CQA services can thus timely inform askers the results so that they do not have to check the progress overtime. 

Nevertheless, the issue is challenging since satisfaction is inherently subjective for askers. It is impractical to directly rank the question-answer pairs relevance since users preferences vary from person to person, although it is the mainstream method in question answering field to recommend best answers. Majority of existing studies in the user satisfaction predictiion adopt feature engineering methods and cast this problem as a binary classification task~\cite{Le2016Evaluating}~\cite{Liu2011Predicting}~\cite{Liu2009Predicting}. They typically employ manually feature selection and apply these features to machine learning algorithms in a supervised manner. Indubitably feature engineering achieves considerable progress, but it is labor intensive and requires cautious design. 

How can we extract and organize discriminative features automatically from data? As the superior performance of deep learning, an intuitive idea is to combine deep learning method to replace manually feature extraction. Moreover, we observe that generally in CQA portals, answers usually come with high diversity but much noise. Users may not assign which answer is the most satisfied, but just close the question as have satisfied with corresponding answers. Under this assumption, we realized that this property actually is applicable to the assumption in multiple instance learning, which indicates that each positive bag must have at least one positive instance. Therefore, we attempt to absorb multiple instance learning into a deep learning framework to assist the task of user personalized satisfaction prediction. Specifically, in our settings, a question with several answers can be treated as a bag with certain instances. We regard a question resolved with at least one satisfactory answer. To this end, we integrate user modeling~\cite{Tang2015User}and recurrent neural network~\cite{hochreiter1997long} with neural tensor network~\cite{socher2013reasoning} to solve the multiple instance learning task, and introduce a \textbf{M}ultiple \textbf{I}stance \textbf{D}eep \textbf{L}earning (\textbf{MIDL}) framework to effectively incorporate users' preferences and the question-answer pairs relevance. The general idea can be illustrated as Figure 1.

\begin{figure}[t]
                        \centering
                        
                        \includegraphics[width=7cm]{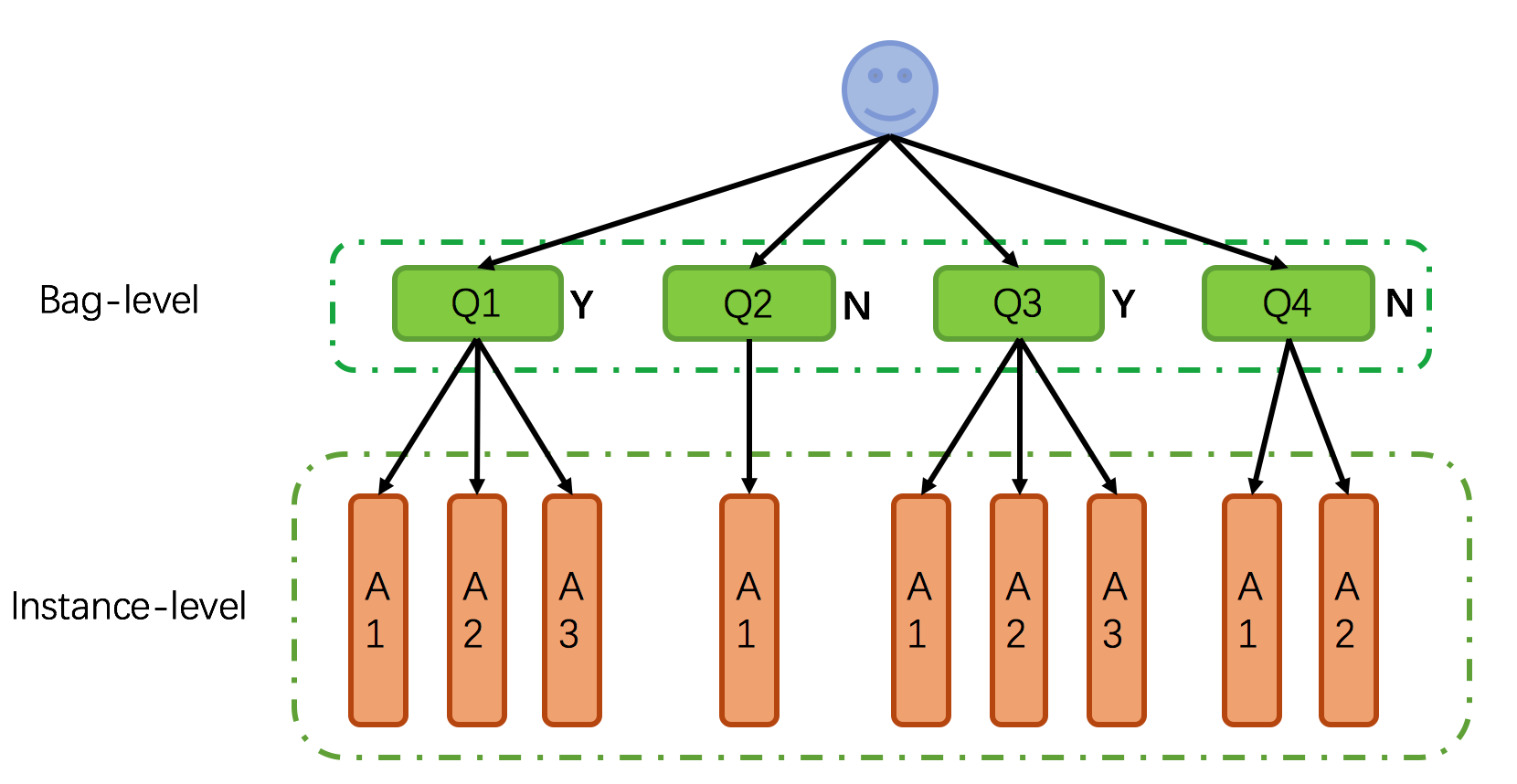}
                        \caption{We denote each question in the bag level and each answer in the instance level. For questions asked by a user, we only know if the question has been assigned satisfied, but we don't know which specific answer is assigned. Here note $Y$ means satisfied and $N$ unsatisfied. }
                    \end{figure}

As is shown, in our settings, we do not need to know which answer will be evaluated as the satisfied one, what we need is the user's reaction to the whole answers on the basis of the question. This can be naturally modeled as multiple instance learning if we consider each answer as an instance and the answers for a question as a bag. 

We conduct experiments to evaluate the effectiveness of the proposed method for the user personalized satisfaction prediction task. The source dataset we process is dumped from StackExchange website. Extensive experimental results show that our assumption of integrating multiple instance learning with deep learning outperforms several strong baseline methods which only use manually feature extraction. Moreover, considering the user's personalized preference shed light on improving effectiveness than just rank the question-answer pairs relevance.

It is worthwhile to highlight several contributions of our work here:
\begin{itemize}
\item We incorporate deep learning into a multiple instance learning framework named \textbf{MIDL} in a principled manner, where we put forward a new assumption in dealing with user personalized satisfaction prediction problem. 

\item Unlike previous studies, our proposed framework which leverages the multiple instance learning assumption and deep learning approach can be processed into an end-to-end procedure. Our framework can be extended into other weakly supervised learning scenarios.

\item Our proposed framework achieves convincing performance than the state-of-the-art models which utilized manually feature extraction. The performance improved significantly in user personalized satisfaction prediction, which demonstrate the potential of our concept of merging multiple instance learning with deep learning.

\end{itemize}

The remainder of this paper is organized as follows. In Section 2, we present a brief view of current related work about user personalized satisfaction prediction and deep learning with multiple instance learning. In Section 3, we formulate the user satisfaction prediction problem and introduce our proposed framework. In Section 4, we describe the experimental settings and report a variety of results to verify the superiority of our model. Finally, we conclude the paper in Section5.

\section{Related Work} 
We briefly review the related work on predicting users personalized satisfaction and the early approaches in studying multiple instance learning as well as current neural tensor network work in this section.
\subsection{Users Personalized Satisfaction Prediction}
Community-based question answering field has attracted substantial researchers to develop various algorithms to better retrieve and extract high-quality relevant information among participants. In previous studies, CQA researchers mainly focus on ranking the answers relevance and diversity, and regard the best ranking answers as the most satisfied results~\cite{fang2016community},~\cite{xumodeling},~\cite{omari2016novelty},~\cite{Andreas2016Learning},~\cite{Shen2015Question}. A significant difference between QA-pairs ranking and users personalized satisfaction prediction is the user's latent preference. From the user's perspective, subjective response to question formulation, related experts recommending, relevant and novel answers taste vary from person to person. User satisfaction researches are popular in information retrieval field but is scarce in CQA field. The most relevant work with our user satisfaction prediction task in CQA field was presented by Liu~\cite{Liu2009Predicting} in 2008. Liu et. directly studied the satisfaction from CQA information seeker perspective, they incorporated a variety of content, structure and community-focused features into a general prediction model. Latha~\cite{Latha2011IMPROVISATION} integrated the available indicators and explored automatic ranking without explictly asking users to assess. In information retrieval field, Liu~\cite{Liu2011Predicting} analyzed unique characteristic of web searcher satisfaction in three aspects: query clarity, query-to-question match, and answer quality. Hassan~\cite{hassan2011task} performed a large scale clickthrough data to explicit judge the user's sequential satisfaction level in the entire search task. Wang~\cite{wang2014modeling} hypothesized that users' latent satisfaction in action-level influences the overall satisfaction and built a latent structural learning method with rich structured features. Liu~\cite{liu2015different} collected the user's feedback with movement data to alleviate some interactions without clickthrough record. We note that these existing methods in predicting the user's satisfaction are mainly depend on artificial extraction characteristics. Although they may gain considerable results, it is too labor intensive. As the flourish of deep learning, it may shed light on this problem. 

\subsection{Multiple Instance Learning}
We observe that most deep learning method are applied in fully supervised settings. However, in our assumption, predicting the user's satisfaction reaction under the condition that each question followed with several unlabeled answers, is basically a weakly supervised problem. In multiple instance learning settings, a bag with several unlabeled instances is assigned positive if and only if it contains at least one positive instance. Since the emergency of multiple instance learning by drug activity prediction researchers in 1990s~\cite{Dietterich1997Solving}, a number of researches have gain significant improvements. For example, Andrew~\cite{Andrews2002SupportVM} introduced MI-SVM and miSVM respectively from the bag-level and the instance-level. Zhang~\cite{Zhang2001EMDDAI} improved DD algorithm by combining EM method and achieved the best result in the musk molecular data at that time. Vezhnevets~\cite{Vezhnevets2010Towards} introduced Semantic Texton Forest to address the task of learning a semantic segmentation using multiple instance learning. Recently, researchers began to incorporate deep representations with multiple instance learning to enhance the performance. Specifically, Wu~\cite{Wu2015DeepMI} designed CNN feature extraction method to jointly exploit the object and annotation proposals in vision tasks including classification and image annotation. Kraus~\cite{Kraus2016ClassifyingAS} also studied a new neural network architecture with multiple instance learning in order to classify and segment microscopy images using only whole image level annotations. Xu~\cite{Wu2015DeepMI} adopted multiple instance learning framework in classification training with deep learning features for medical image analysis. Zhou~\cite{Zhou2004Multi} investigated the web index recommendation problem from a multiple intance view, they regarded the whole website as a bag and the linkpages in website as the corresponding instances. We note that in multiple instance learning field, rare researchers have exploit deep learning tactics into Natural Language Processing task. Thus we further attempt to extend the application into CQA field.

\subsection{Neural Tensor Network}
Previous models suffer from weak interaction between two entities in the vector space. To address this problem, Socher~\cite{socher2013reasoning} first introduced the neural tensor network to allow the entities and relations to interact multiplicatively. They successively applied the neural tensor network to solve the problem in typical Natural Language Processing field.~\cite{socher2013reasoning} focused on predicting additional true relations between entities.~\cite{chen2013learning} studied the problem of learning new facts with semantic words.~\cite{Socher2013Recursive} introduced a recursive neural tensor network to remedy sentiment detection task. Neural tensor network out-performed other linear combination approaches significantly and raised much attention among researchers. In CQA field, researchers also adopt the idea of neural tensor network. Xia~\cite{xumodeling} modeled document novelty with neural tensor network for search result diversification task, they automatically learned a nonlinear novelty function based on preliminary representations of a document and other candidate documents. Qiu~\cite{Qiu2015Convolutional} integrated Q-A pairs semantic matching with convolutional and pooling layers, and exploited neural tensor network to learn the matching metrics. In our paper, we integrate neural tensor network to link the relevance of the user's attitude towards to the question accompanied with answers.

\section{Multiple Instance Deep Learning}
In this section, we propose the framework of Multiple Instance Deep Learning~\textbf(MIDL). We first introduce the task that we are seeking to solve and frame our formulation. Then we present the details of learning textual contents of U-Q-A representations with Recurrent Neural Network. And then we provide conceptual setting of multiple instance learning with neural tensor network. Finally we introduce the training process and corresponding algorithm.

\subsection{Task Description and Formulation}
In this paper, we focus on predicting users personalized satisfaction. As is described earlier, in our formulation, we aware that it is reasonable to formulate that a user's satisfaction reaction lies in at least one of the corresponding satisfactory answers. In real world, when faced with a list of answers, users may have difficulties in deciding which answers are satisfied. However, it is justifiable to assume that the questions resolved with at least one satisfactory answer. In other words, it is natural to treat a question resolved as a positive bag with at least one of positive satisfactory answer instances. This property inspires us to design a multiple instance learning tactic to model the satisfaction prediction task.

Detailed manual annotations for each answer are time consuming for QA users. An alternative is to learn the global annotations for the overall answers, which is the main idea of multiple instance learning. Given the multiple instance learning assumption, questions with corresponding answers are organized as bags, which denotes as $\left \{ \chi_{i}\right \}$. Within each bag there are a set of answer instances $\left \{ \chi _{ij} \right \}$. We define the users satisfaction reactions as the labels $\left \{ Y _{i} \right \}=\left \{ 1,-1 \right \}$. In our proposal, The labels $\left \{ Y _{i} \right \}$ are only available at the bag level, and we do not know the label at the instance level $\left \{ y_{ij} \right \}$. The task is to predict the labels of unseen bags with multiple instances. We thus incorporate the multiple instance learning property into predicting the label of the user's satisfaction reaction at the bag level.

\subsection{Modeling U-Q-A with Recurrent Neural Network}
Considering the flourish of deep learning and the ideas of learning from data, an intuitive method is to combine deep learning method to replace manually feature extraction in learning the semantic embedding of questions and answers textual contents. In~\textbf{MIDL} framework, we exploit Bi-directional LSTM for learning Q-A deep representations, which is inspired by~\cite{Melamud2016context2vecLG}. The structure of our proposed Bi-directional LSTM is shown in Figure2.
\begin{figure}[t]
                        \centering
                        
                        \includegraphics[width=8cm]{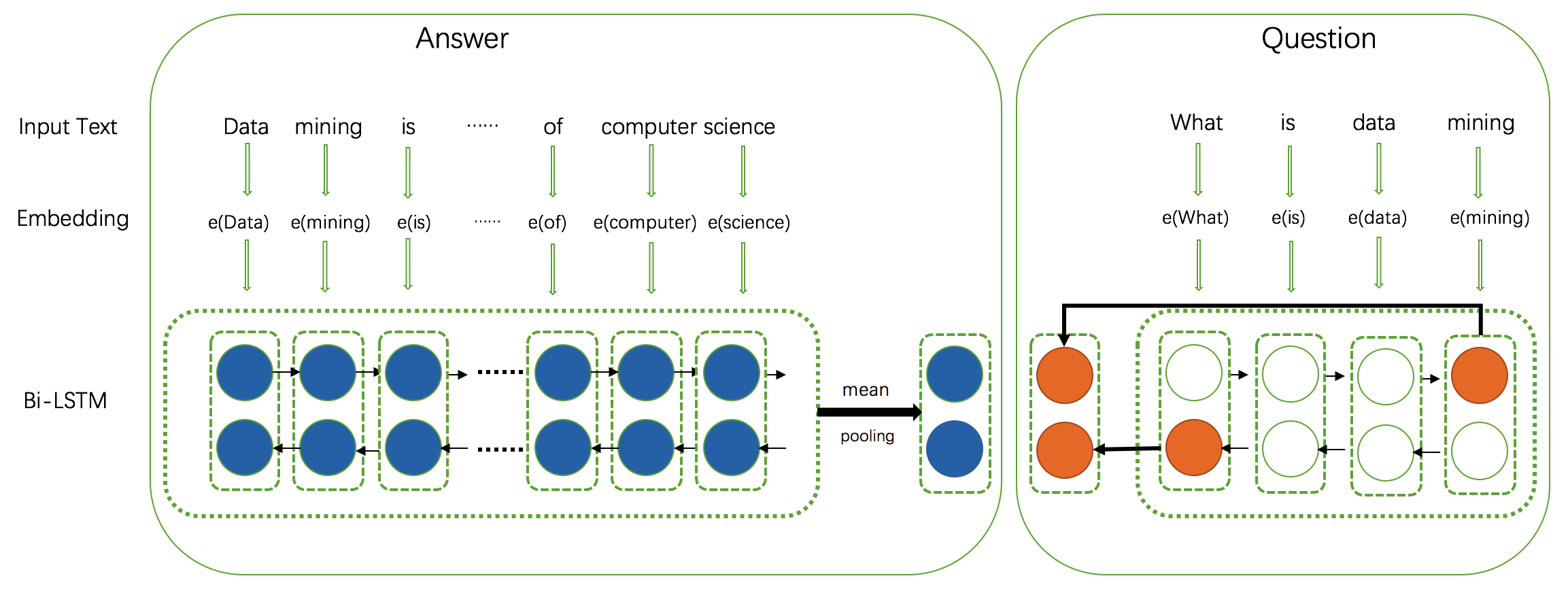}
                        \caption{We adopt one word embedding function and two encoders to encode answers and questions with Bi-directional LSTM respectively. For answer encoder, we concatenate each word within two layers and assign a mean pooling to get the global embedding of the answer. For question encoder, we simply concatenate the last hidden state in both layer to get the question embedding. }
                    \end{figure}

Intuitively, our framework of modeling U-Q-A semantic embedding is structured as follows:
\begin{enumerate}
\item We define the semantic embedding of the user in the common user space.            
\item We apply a question encoder and an answer encoder, which compute each individual word of the contents into contextual vector embedding with Bi-directional LSTM.
\item We concatenate the user embedding with question embedding and obtain the new semantic vectors of Q-U embedding. 
\end{enumerate}

In detail, we employ one word embedding encoder and two Bi-directional LSTM encoders to encode User-specific-Question embedding and Answer embedding into hidden vectors. We believe that using Bi-directional LSTM can better capture the contextual information as it can reduce the vanishing gradient problem. A Bi-directional LSTM consists of a forward LSTM and a backward LSTM. The forward LSTM reads each word $w_i$ (i.e., from $w_{1}$ to $w_{i}$) in sequence as it is ordered, and generate the hidden states of each word as $\left ( \overrightarrow{h_1},...,\overrightarrow{h_i} \right )$. For the backward LSTM, it processes each sentence in its reversed order (i.e., from $w_{i}$ to $w_{1}$) and form a sequence of hidden states $\left ( \overleftarrow{h_1},...,\overleftarrow{h_i} \right )$. We calculate the hidden states $\overrightarrow{h_i}$ by following equations:
\begin{align*}
 i_{t}&=\delta (W_{i}x_{t}+G_{i}h_{t-1}+b_{i}) \\
\hat{C}_{t}&=tanh(X_cx_t+G_fh_{t-1}+b_f) \\
f_t&=\delta (W_fx_t+G_fh_{h-1}+b_f) \\
C_t&=i_t\cdot \hat{C_t}+f_t\cdot C_t\\
o_t&=\delta(W_ox_t+G_oh_{t-1}+V_oC_t+b_o)\\ 
h_t&=o_t \cdot tanh(C_t)
\end{align*}

where $\sigma$ represents the sigmoid activation function; $W_s$, $U_s$ and $V_o$ are weight matrices; and $b_s$ are bias vectors. There are three different gates (input, output, forget gates) for controlling memory cells and their visibility. The input gate can allow incoming signal to update the state of the memory cell or block it and the output gate can allow the state of the memory cell to have an effect on other neurons or prevent it. Moreover, the forget gate decides what information is going to be thrown away from the cell state. We take the output of the last LSTM cell, $h_k$, as the semantic embedding of the input sequence $\left \{x_1,x_2,...,x_k \right \}$.

Specifically, we design two practices of Bi-directional LSTM methods. For question encoder, we concatenate the forward and backward last hidden states from respective recurrent networks, and denote $\left \{ h_{x,i}=\left [ \overrightarrow{h_i},\overleftarrow{h_1} \right ] \right \}$ as the question semantic embedding $f_i(q)$. For answer encoder, since we care more about the relevance of each word in answers corresponds to the question, we encode every word contextual embedding from the context of answers, and denote $\left \{ h_{x,i}=\left [ \overrightarrow{h_i},\overleftarrow{h_i} \right ] \right \}$ as the answer semantic embedding $f_i(a)$. And then we put a mean pooling layer to obtain the general semantic embedding of each answer. Both encoders use the same word embedding as the input. In our models, we implement the word embedding function in a usual way, which exploit a look-up table and each word is indexed by one-hot representation from the vocabulary.

\subsection{Exploiting Multiple Instance Learning with Neural Tensor network}
To model the user's attitude towards to the answers, we propose to use neural tensor network to measure the relationships between Q-U representation and the answers representations. Neural tensor network is proposed for reasoning over relationships between two entities~\cite{socher2013reasoning}. Given two entities $(e_1,e_2)$ encoded with $d$ dimension, we use neural tensor network to state whether these two entities have a certain relationship $R$, and what the certainty is. We adopt the neural tensor network with a bilinear tensor layer to compute the relevance of two entity vectors across multiple dimensions. Assume $e_1,e_2\in \mathbb{R} ^{d}$ is the vector representations of the two entities, we compute the score of these two entities in a certain relationship. The equation is presented in the following:

\begin{equation}
g(e_1,R,e_2)=\mu _{R}^{T}tanh(e_1^TW_R^{[1:z]}e_2+V_R\begin{bmatrix}
e_1\\ 
e_2
\end{bmatrix}+b_R) 
\end{equation}

where $W_R^{[1:z]}\in \mathbb{R}^{d\times d\times z}$ is a tensor and we conduct the bilinear tensor product $e_1^TW_R^{[1:z]}e_2$ to gain a vector $h\in \mathbb{R}^d$. Each entry of $h$ is computed by one slice $i=1,...,z$ of the tensor: $h_i=e_1^TW_R^{[1:z]}e_2$. The other parameters for relation $R$ are the standard form of neural network: $V_R\in \mathbb{R}^{z\times 2d}$ and $U\in \mathbb{R}^z$, $b_R\in \mathbb{R}^z$. We reveal the original neural tensor network in Figure3.

\begin{figure}[t]
                        \centering
                        
                        \includegraphics[width=8cm]{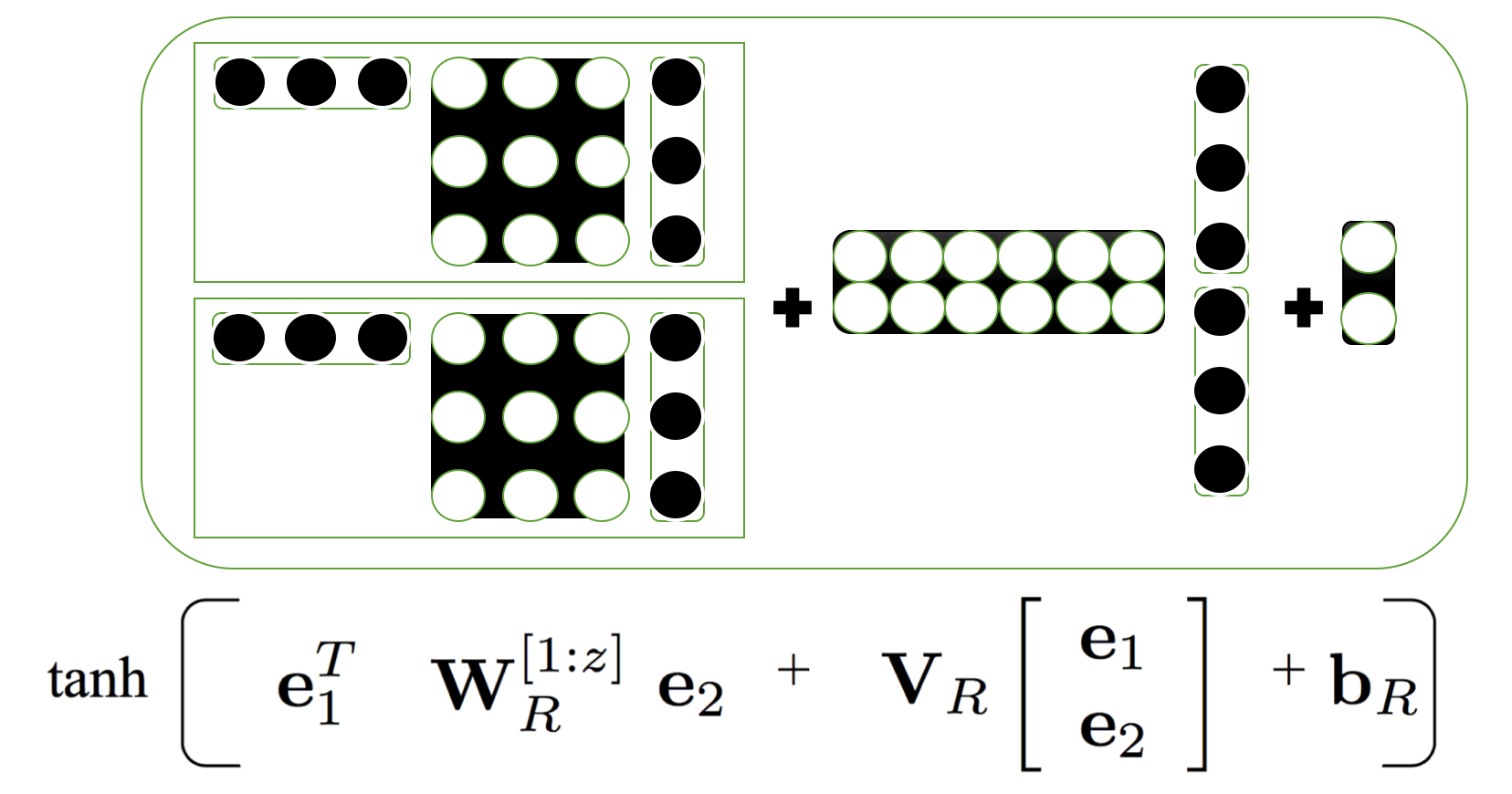}
                        \caption{Visualization of the neural tensor network applied for entities relationships measurement. }
                    \end{figure}

\begin{figure*}[!htbp]
                        \centering
                        
                        \includegraphics[width=17cm]{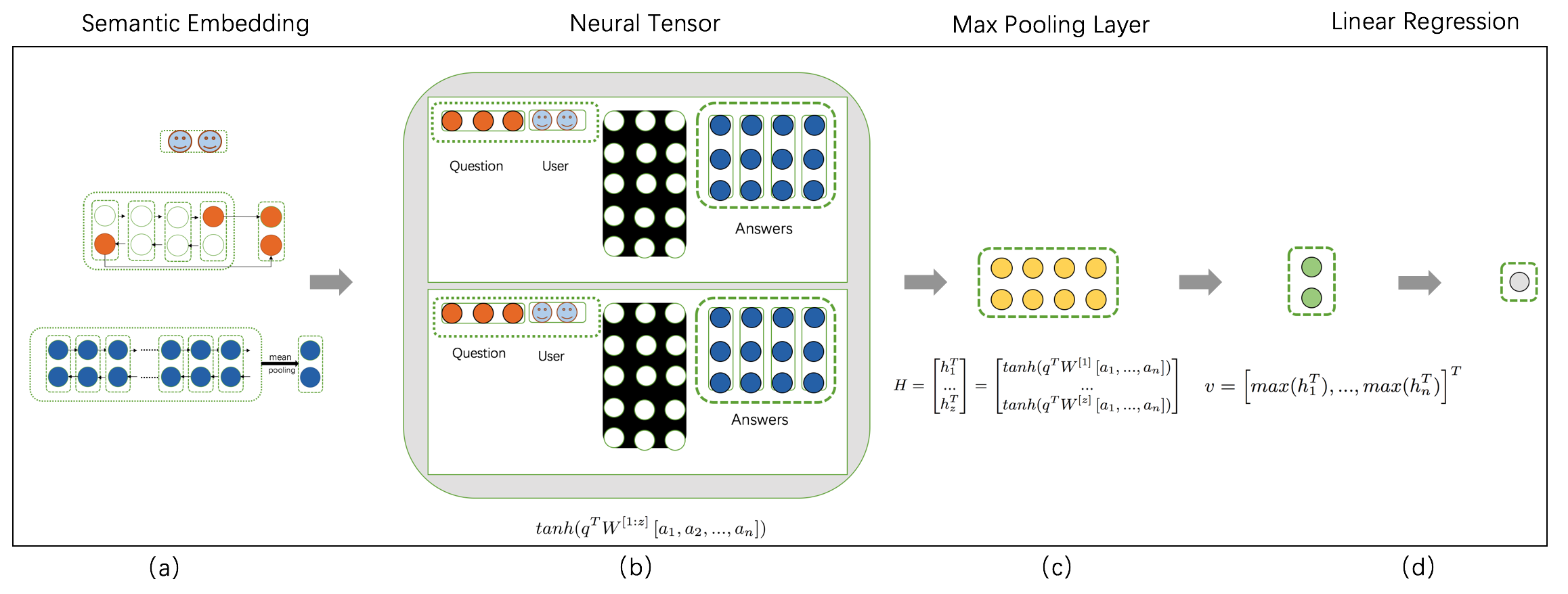}
                        \caption{ The overview of our proposed framework MIDL. (a)We adopt Bi-directional LSTM to learn the contextual content embedding of questions and answers, and initialize the user vector. (b) We concatenate the question embedding with user vector to form a Q-U representation. We then utilize the modified neural tensor network to model the relationships of Q-U representation and corresponding answers. (c)We put a max pooling layer to extract the most representative element, the pooling result represent the whole bag embedding for multiple instance learning. (d) The bag-level vectors are applied into logistic regression and obtain the predicting result of user satisfaction.}
                    \end{figure*}

Intuitively, the origin neural tensor network is proposed to model the relationships between two entities with a bilinear tensor product. This conception can be naturally extended into modeling the relationships of a Q-U representation with respect to the answers representations. To this end, we adopt the neural tensor network into our multiple instance learning framework. The schematic diagram of our proposed framework is shown in Figure4.

To learn multiple instances as a bag of samples, we incorporate the Q-U-A deep representations with multiple instance learning. We apply the modified version of neural tensor network to jointly learn the multiple instances within a bag. More specifically, assume that given the Q-U embedding $Q=\left \{d_i  \right \}$ and the set of n answer embedding $A=\left \{d_j \right \}_{j=1}^n$. All of the embedding are obtained from preliminary Bi-directional LSTM representation. Given a Q-U representation $q \in Q$ and a set of answers $\left \{a_1,a_2,...,a_n \right \}$. We extend the origin neural tensor network in the following equation:

\begin{equation}
g_n(q,A)=\mu ^T max \left \{ tanh(q^TW^{[1:z]}\left [ a_1,a_2,...,a_n \right ])\right \}
\end{equation}

We define the answers preliminary representation vectors $\left [ a_1,a_2,...,a_n \right ] $ and form a matrix $M \in \mathbb{R}^{d \times n}$. $W_R^{[1:z]}\in \mathbb{R}^{d\times d\times z}$ is a tensor. For convenience we ignore the other bias term in original neural tensor network. We also conduct the bilinear tensor product $q^TW_R^{[1:z]}\left [ a_1,a_2,...,a_n \right ]$ followed by a nonlinear operation:

\begin{equation}
H=\begin{bmatrix}
h_1^T\\ 
...\\ 
h_z^T
\end{bmatrix}
=\begin{bmatrix}
tanh(q^TW^{\left [ 1 \right ]}\left [ a_1,...,a_n \right ])\\ 
...\\ 
tanh(q^TW^{\left [ z \right ]}\left [ a_1,...,a_n \right ])
\end{bmatrix}
\end{equation}

Where $h_i \in \mathbb{R}^n$ is achieved by each slice of the tensor. Here we apply the hidden states $H$ to model the user personalized attitude towards to a question with corresponding answers. The output of $H$ is a matrix $z\times n$, in which each column is the representation of an instance. We aggregate the representation of the bag for multiple instance learning with max pooling: 

\begin{equation}
v=\left [ max(h_1^T),...,max(h_n^T) \right ]^T
\end{equation}

Here we use max-pooling to extract the most significant element to well represent the whole bag for multiple instance learning. And finally, we adopt the bag representation $v$ into a binary logistic regression which denotes ``satisfied'' or ``not satisfied'' to predict the label of the bag. Specifically, we formulate a binary multiple instance learning framework which optimized the loss function of bag classification. Denote $X_i=\left \{ X_{i1},X_{i2},...,X_{im} \right \}$ is the $i^{th}$ bag of the question in the training set, and $\left \{ X_{i1},X_{i2},...,X_{im} \right \}$ is the answer instances. $m$ is the number of answer instances in the $i^{th}$ bag. $y_i\in \left \{ -1,+1\right \}$ is the label of the bag, 1 denotes ``satisfied'' and -1 denotes ``not satisfied''. The loss function is:

\begin{equation}
L(H)=-\sum_{i=1}^{n}\textbf{1}(y_i=1)logH(X_i)+\textbf{1}(y_i=-1)log(1-H(X_i))
\end{equation}

Where $\textbf{1}\left ( \cdot  \right )$ is an indicator function.

We iteratively train weak classifiers ${h}'(x)$ using gradient descent:

\begin{equation}
w_{ij}=\frac{\partial L(H)}{\partial h(x_{ij})}=-\frac{\partial L(H)}{\partial H(X_i)}\frac{\partial H(X_i)}{\partial h(x_{ij})}
\end{equation}

where $h(x)$ updates by $h(x)+\alpha {h}'(x)$ and $\alpha$ is the parameter optimized by line searching. So far we generate an efficient classifier after the loss function converge.

\subsection{Training}
In this section, we present the details of our multiple instance deep learning \textbf{MIDL} method and summarize the main training process in Algorithm 1.

\begin{algorithm}
\caption{\textbf{MIDL for Users Satisfaction Prediction}}
\begin{algorithmic}[1]
\REQUIRE ~~\\ 
Input: Question-Answer Dataset $D(Q,A,U_{id})$, \\question q, askerid uid, the $i_{th}$ answers set of q is $A\_q$  
\STATE Pre-train the word-embedding of $Q$ and $A$ by skip-gram
\STATE Initialize the user embedding
\FOR{$q$ in $Q$} 
   \FOR{$a$ in $A\_q$} 
        \STATE$a\_emb=lstm(a)$
   \ENDFOR
   \STATE$q\_emb=lstm(q)$     
   \STATE$u\_emb=U(uid)$
   \STATE neural-tensor($q\_emb$,$a\_emb$,$u\_emb$)
   \STATE Summate the total training loss
   \STATE Update parameters by SGD
\ENDFOR
\end{algorithmic}
\end{algorithm}

We begin with one-hot representations on each word, then we apply two Bi-directional encoders to denote questions and answers semantic representations respectively and we initialize the user embedding. After that, we concatenate each question with its asker to form the Q-U representation, which represents the asker's intent to the question. Afterwards we apply the Q-U representation with a set of answers to the updated neural tensor network to compute the relationships. And finally we use the logistic regression to predict the satisfaction level of users.

Denote all the parameters in our framework as $\Theta$, we define the objective function in training process:

\begin{equation}
\underset{min}{\Theta}L(\Theta)=L(\Theta)+\lambda \left \| \Theta \right \|_{2}^{2}
\end{equation}

$\lambda>0$ is a hyper-parameter to trade-off the training loss and regularization. By using SGD optimization, at time step t, the parameter $\Theta$ is updated as follows:
\begin{equation}
\Theta_t=\Theta_{t-1} - \frac{\rho }{\sqrt{\sum_{i=1}^{t}g_i^2}}g_t
\end{equation}

where $\rho$ is the initial learning rate and $g_t$ is the subgradient at time t.

\section{Experiments}
To empirically evaluate and validate our proposed framework multiple instance deep learning\textbf{(MIDL)}, we conduct experiments on a widely used dataset dumped from Stack Exchange community.

\subsection{Data Preparation}
 The dataset downloaded from the famous community-based question answering portal Stack Exchange is an anonymized dump of all user-contributed content. The whole dataset consists of over 133 question answering forums and the StackOverFlow is the biggest forum among them. In our experiment, we snapshot four forums history data to validate our framework against some baselines. The theme of these four forums are ``Android'', ``Academia'', ``Photo'', ``Christian''. We present the detail of these four forums data in Table1.

\begin{table}[htbp]
 \caption{\label{tab:test}Statistic of the four forums data}
 \centering
 \begin{tabular}{|c|c|c|c|c|}
 \hline
  Forum    & Question  & Answer  &User   &Satisfied\\  \hline
  Android  & 25310     &42238   &15845   &42.1\%  \\ \hline
  Academia & 12062     &31046    &5875    &50.6\%  \\ \hline
  Photo    & 14414     &38206    &6867    &59.6\% \\ \hline
  Christian & 6915     &17502    &1777    &53.9\%\\ \hline
 \end{tabular}
 \end{table}

As we can see, questions in four forums received distinct proportion of answers, and the average satisfied ratio vary from each other. Among these four forums, the Android forum is the most popular but draws on only 1.67 answers for each question on average and the user's satisfaction level is the lowest compared with other forums. The Photo forum get the highest satisfaction level with 59.6\% and the most answers with 2.65 answers per question. In summary, asker satisfaction and other statistics of the questions vary widely from each forum data. We then split the dataset into training set, validation set and testing set without overlapping in our experiments. We fix the validation set as 10\% of the total data to tune the hyperparameters and the size of testing set is 30\%.

\subsection{Evaluation Criteria}
In order to evaluate the performance of different models, we employ Precision, Recall, F1-Measure and Accuracy as evaluation measures. These measure criterions are widely used in the evaluation for user satisfaction prediction task. Precision reports the ratio of the predicted satisfied question respect to the indeed rated satisfactory by users. Recall evaluates the fraction of all the indeed rated satisfactory questions that are distinguished by the framework. F1-Measure comprehensive analysis the results of Precision and Recall. Accuracy reflects the framework classification ability for the entire sample.

We first denote four indicators to signify the derivations and the Terminology. We define \textbf{TP} as the true positive, \textbf{TN} as the true negative, \textbf{FP} as the false positive, \textbf{FN} as the false negative.

\begin{itemize}
\item \textbf {Precision} takes all the retrieved documents into account, it measures the fraction of retrieved documents that are relevant to the query in the field of information retrieval.
\begin{equation*}
Precision=\frac{TP}{TP+FP}
\end{equation*}

\item \textbf {Recall} can be regarded as the probability that a relevant document is retrieved by the query, it measures the fraction of the documents that are relevant to the query that are successfully retrieved.
\begin{equation*}
Recall=\frac{TP}{TP+FN}
\end{equation*}

\item \textbf {F1-Measure} compromises the bias of Precision and Recall, it seeks for balance between Precision and Recall with evenly weighted parameter so that can be criticized in particular circumstances.

\begin{equation*}
F1=2\times \frac{Precision \cdot Recall}{Precision+Recall}
\end{equation*}

\item \textbf {Accuracy} measures the fraction of the true classified documents that are relevant to the whole sample.
\begin{equation*}
Accuracy=\frac{TP+TN}{TP+TN+FP+FN}
\end{equation*}
\end{itemize}

\subsection{Performance Comparisons}
To validate the performance of our approach, we compare our proposed method against with other eight state-of-the-art methods for the users personalized satisfaction prediction problem.  

\begin{itemize}
\item \textbf{ASP\_SVM} Support vector machines with manually selected features in~\cite{Liu2009Predicting}. SVM proposed by Vapnik ~\cite{Cortes1995Support} is considered as a strong supervised learning algorithm that analyzes data used for classification task. In our experiment, we implement the relevant feature selection according to illustration in ~\cite{Liu2009Predicting}, and then we use libsvm to integrate the features to svm to classify the label of the user's satisfaction result.

\item \textbf{ASP\_RF} RandomForest with manually selected features in~\cite{Liu2009Predicting}. Random forests are an ensemble method which was created by TK~\cite{Ho1995Random}. We use random forest classifier as well as feature selection in order to get high precision on the target label.  

\item \textbf{ASP\_C4.5} C4.5 algorithm with manually selected features in~\cite{Liu2009Predicting}. C4.5 is used to generate a decision tree developed by JR Quinlan~\cite{Quinlan1993C4}, and has become quite popular in classification. Here we use the same feature selection referred to ~\cite{Liu2009Predicting}.  

\item \textbf{ASP\_Boost} Boosting algorithm with manually selected features in~\cite{Liu2009Predicting}. Boosting posed by Kearns~\cite{Kearns1989Crytographic} is primarily applied to reduce bias and variance in supervised learning, the idea of boosting is also from ensemble methodology.

\item \textbf{ASP\_NB} Naive Bayes with manually selected features in~\cite{Liu2009Predicting}. Naive Bayes classifier is based on applying Bayes' theorem with strong independence assumptions between the features, in this paper we also conduct the Naive Bayes classifier with selected features to fully evaluate the feasibility of our framework.

\item \textbf{MISVM} MISVM Proposed by Andrews~\cite{Andrews2002Support} is a classical multiple instance learning algorithm, it extend SVM to maximize the bag-level pattern margin over the hidden label variables. Here we address the predicting problem with MISVM to suit our settings.  

\item \textbf{EM-DD} Em-DD is a general-purpose for multiple instance problem that combines EM with the diverse density(DD) algorithm~\cite{Zhang2001EMDDAI}. We derive the idea of EM-DD algorithm to compare the performance with MIDL framework.

\item \textbf{BP-MIP} BP-MIP~\cite{zhou2002neural} employs a specific error function derived from BP neural network. We implement the simplified version of BP-MIP to address our problem. 
       
\end{itemize}

Overall, the first five classification baselines are supervised methods which focus on the feature selection manner and latter three are weakly supervised methods which applied in multiple instance learning. In order to better demonstrate the impact of different components of our proposed framework \textbf{MIDL}, we respectively evaluate the performance between manually feature selection and deep learning representations, and validate the feasibility of our assumption with typical multiple instance learning algorithms. In our experiments, we select the available features according to the reference of the paper~\cite{Liu2009Predicting}. we totally organized five basic entities in question answering community, which is questions, answers, Q-A pairs, users and categories. And we implement five representative algorithms of the state-of-the-art in classification. The hyperparameters and parameters which achieve the best
performance on the validation set are chosen to conduct the testing evaluation.  

\subsection{Experimental Results and Analysis}
To evaluate the performance of our proposed framework, we conduct several experiments on four metrics described above.

Table 1, 2, 3 and 4 show the evaluation results on Precision, Recall, F1-Measure and Accuracy, respectively. We conduct the experiments with four datasets extracted from Stack Exchange website. We choose the parameters which achieve the best performance to implement the testing evaluation. We then report several interesting analysis that we observed on the evaluation results .  

\begin{table}[htbp]
 \caption{\label{tab:test}Experimental results on Precision with different community datasets for training.(best scores are boldfaced)}
 \centering
 \begin{tabular}{lclccccl}
  \toprule

  Dataset& Android & Academia & Photo & Christian \\
  \midrule
 ASP\_SVM   &0.7979 &0.8054 &0.8195 &0.7963  \\       
 ASP\_RF    &0.8031 &0.8265 &0.8044 &0.8187 \\      
 ASP\_C4.5  &0.8002 &0.8337 &0.8025 &0.7846 \\      
 ASP\_Boost &0.7969 &0.8271 &0.8039 &0.8143 \\       
 ASP\_NB    &0.7633 &0.7835 &0.7154 &0.7965 \\       
 MISVM      &0.7201 &0.7743 &0.7982 &0.7644 \\
 EM-DD      &0.7531 &0.7557 &0.7879 &0.7212 \\
 BP-MIP     &0.7748 &0.8153 &0.7294 &0.7238 \\
 MIDL     &\textbf{0.8113}   &\textbf{0.8744} &\textbf{0.8563} &\textbf{0.8195}\\    

  \bottomrule

 \end{tabular}
\end{table}

\begin{table}[htbp]
 \caption{\label{tab:test}Experimental results on Recall with different community datasets for training.(best scores are boldfaced)}
 \centering
 \begin{tabular}{lclccccl}
  \toprule

Dataset& Android & Academia & Photo & Christian \\
  \midrule
 ASP\_SVM   &0.7544 &0.7458 &0.8003 &0.7914  \\       
 ASP\_RF    &0.8612 &\textbf{0.8361} &0.8014 &0.8117 \\      
 ASP\_C4.5  &0.7886 &0.8053 &0.8152 &0.8089 \\      
 ASP\_Boost &0.7935 &0.7731 &0.7841 &0.7578 \\       
 ASP\_NB    &0.7917 &0.7659 &0.7452 &0.7335 \\       
 MISVM      &0.7382 &0.7115 &0.8674 &0.7361 \\
 EM-DD      &0.7964 &0.7411 &0.7238 &0.7525 \\
 BP-MIP     &0.8192 &0.7871 &0.7493 &0.8716 \\
 MIDL     &\textbf{0.9773} &0.7966 &\textbf{0.8947} &\textbf{0.9222}\\    
 \bottomrule

 \end{tabular}
\end{table}

\begin{table}[htbp]
 \caption{\label{tab:test}Experimental results on F1-Measure with different community datasets for training.(best scores are boldfaced)}
 \centering
 \begin{tabular}{lclccccl}
  \toprule

  Dataset& Android & Academia & Photo & Christian \\
  \midrule
 ASP\_SVM   &0.7755 &0.7746 &0.8098 &0.7938  \\       
 ASP\_RF    &0.8311 &0.8313 &0.8029 &0.8152 \\      
 ASP\_C4.5  &0.7944 &0.8193 &0.8088 &0.7966 \\      
 ASP\_Boost &0.7952 &0.7992 &0.7939 &0.7850 \\       
 ASP\_NB    &0.7772 &0.7746 &0.7300 &0.7637 \\       
 MISVM      &0.7290 &0.7416 &0.8314 &0.7500 \\
 EM-DD      &0.7741 &0.7483 &0.7545 &0.7365 \\
 BP-MIP     &0.7964 &0.8010 &0.7393 &0.7909 \\
 MIDL     &\textbf{0.8866} &\textbf{0.8337} &\textbf{0.8751} &\textbf{0.8678}\\    

  \bottomrule

 \end{tabular}
\end{table}

\begin{table}[htbp]
 \caption{\label{tab:test}Experimental results on Accuracy with different community datasets for training.(best scores are boldfaced)}
 \centering
 \begin{tabular}{lclccccl}
  \toprule

  Dataset& Android & Academia & Photo & Christian \\
  \midrule
 ASP\_SVM   &0.8091 &0.8013 &0.7996 &0.8146  \\       
 ASP\_RF    &0.8253 &0.8099 &0.8000 &0.8501 \\      
 ASP\_C4.5  &0.8032 &0.7765 &0.7842 &0.8223 \\      
 ASP\_Boost &0.7953 &0.8331 &0.8089 &0.8347 \\       
 ASP\_NB    &0.7554 &0.7883 &0.7969 &0.7839 \\       
 MISVM      &0.7555 &0.7828 &0.7635 &0.7947 \\
 EM-DD      &0.7742 &0.8038 &0.7803 &0.8092 \\
 BP-MIP     &0.7798 &0.8275 &0.7934 &0.8251 \\
 MIDL     &\textbf{0.8429} &\textbf{0.8563} &\textbf{0.8337} &\textbf{0.8901}  \\    
  \bottomrule

 \end{tabular}
\end{table}

As mentioned previously, we argue that users personalized satisfaction can be assumed as a multiple instance learning problem. In order to verify our hypothesis, we conduct eight baselines trained with the same dataset and tested with the same evaluation criteria. The first five baselines utilized the manually selected features introduce in~\cite{Liu2009Predicting}, we strictly follow the instruction of the paper and applied the features into five popular classification algorithm. The latter three baselines are designed to compare the feasibility of our proposed multiple instance framework, we refer to the illustration of these three models and applied our assumption with user satisfaction prediction. Table 2, 3, 4 and 5 show the evaluation results in terms of four typical evaluation criterias Precision, Recall, F1-Measure and Accuracy. Figure 5 explore the tendency of performance with varying amount of training data in our framework. Figure 6 shows the prediction accuracy for distinct groups of users with different number of questions. With these experimental results, we can summarize several interesting points: 
\begin{figure}[t]
                        \centering
                        
                        \includegraphics[width=7cm]{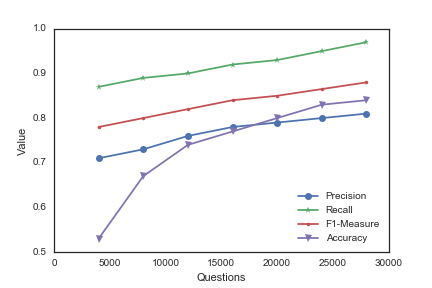}
                        \caption{Precision, Recall, F1-Measure, Accuracy for varying amount of training data in Android CQA forum. }
                    \end{figure}

\begin{figure}[t]
                        \centering
                        
                        \includegraphics[width=7cm]{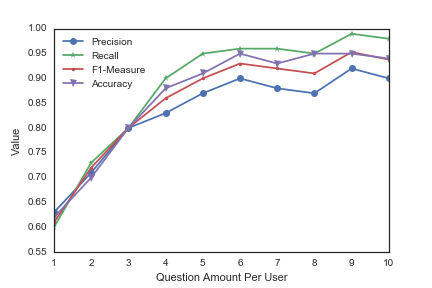}
                        \caption{Precision, Recall, F1-Measure, Accuracy for varying active level of users, here we use average questions per user as the group clustering criteria in Android CQA forum. }
                    \end{figure}

\begin{itemize}
\item We observe that almost in four forum datasets our proposed framework MIDL outperforms other baselines significantly, which suggests that it is feasible for us to hypothesis the users personalized satisfaction prediction problem is appropriate for multiple instance learning. 
\item We notice that under the same selected features, the ensemble approaches like Random Forest, C4.5 and Boosting algorithms achieved better performance than single statistic based classification algorithms.
\item Compared with artificial feature selection models, our framework MIDL gains better experimental results. moreover, our framework is easier to train with deep learning tactic.
\item We implement three typical multiple instance learning algorithms MISVM, EM-DD and BP-MIP. These three algorithms achieved superior performances in their own problem settings. However, in predicting the user's satisfaction towards to bags of answers, they do not work well. We conjecture that this is due to the problem settings and obviously our framework is more appropriate for the user satisfaction prediction task.
\item It is no surprising to see from Figure 5 that with sufficient training data, we can achieve a better performance since deep learning method can learn more accurate representations from the big data.
\item The accuracy increased with more records for individuals. From Figure 6 we notice that the prediction dramatically increases for users with varying amount of questions. The tendency of the folding lines arise as the number of questions per user increases. And we can clearly see that the folding lines slow down and tend to consistant after reaching 5 quesitons per user. So we can conclude that if we want to obtain a better prediction results, we need at least 5 records for per user.

\end{itemize}

\section{Conclusion}
Users satisfaction prediction is an essential component in Community Question Answering(CQA) services. Existing approaches have been hurt from the necessaries of predefining artificial selected features, which are usually difficult to design and labor-intensive in real applications. In this paper we formulate the user satisfaction prediction problem as a multiple instance learning pattern, and discuss a new framework which is capable of exploiting deep learning representations associated with our assumption to enhance the weakly supervised learning ability. We develop a neural tensor network based method with Bi-directional LSTM for evaluating the user's attitude towards a set of answers related to the proposed question. Our approach can be applied easily to existing information retrieval models and extended into other user satisfaction modeling field. Experimental results conducted on a large CQA data set from Stack Exchange demonstrate the significant improvement of the proposed technique.

This work opens to several interesting directions for future work. First, it is of relevance to apply the proposed technique to other information retrieval approaches. What's more, we can use more complex means to model the users latent preference and enhance the performance. Moreover, applying multiple instance learning with deep learning tactic into Natural Language Processing field is a big treasure to hunt. As future work, we will extend the multiple instance learning assumption into more applicable scenarios.

\bibliographystyle{abbrv}
\bibliography{sigproc}  

\end{document}